# New constraints for CP-violating forces between nucleons in the range $10^{-4}$ cm ÷ 1 cm

A.P. Serebrov

*Petersburg Nuclear Physics Institute RAS*

**Abstract**

The range of nucleon interaction $10^{-4}$ cm ÷ 1 cm is interesting because it corresponds to the mass range of a intermediate particle inside so named "axion window" that is not closed yet by experiment. Depolarization of ultracold neutrons (UCN) during their storage in material traps can be caused by CP-violating pseudo-magnetic precession of the neutron spin in the vicinity of the unpolarized substance surface. Using the experimental limits for UCN depolarization new constraints were set for the product of the scalar, pseudo-scalar dimensionless constants $g_S g_P$ and the parameter $\lambda_{PS} = \hbar / m_{PS} c$, determining the Yukawa-type of the nucleon interaction potential via new pseudo-scalar boson (axion-like particle) with a mass of $m_{PS}$:

$$g_S g_P \lambda_{PS}^2 \leq 2.96 \cdot 10^{-21} \ [\text{cm}^2] \qquad \text{for } 10^{-3} \text{ cm} < \lambda_{PS} < 1 \text{ cm};$$

$$g_S g_P \lambda_{PS}^2 \leq 3.9 \cdot 10^{-22} \ [\text{cm}^2] \qquad \text{for } 10^{-4} \text{ cm} < \lambda_{PS} < 10^{-3} \text{ cm}.$$

Improvement of the limit for $g_S g_P$ in the area of $\lambda_{PS}$ from 0.1 cm to 1 cm accounts for 4÷5 orders of magnitude in comparision with previous limit.

The prospects of increasing in accuracy search for CP-violating pseudo-magnetic precession are considered. The estimations of the possible effects of pseudo-magnetic precession in the frame of the theoretical models with CP-violation are discussed.

The work [1] suggests searching for the pseudo-magnetic neutron spin precession in the vicinity of the substance surface by measuring the resonance shift in a magnetic resonance spectrometer with ultracold neutrons. This work demonstrates that due to the pseudo-magnetic precession besides the neutron resonance shift also occurs the effect of UCN depolarization during their storage in material traps. Using available experimental data on UCN depolarization from the works [2-4] we can set new constraints for the parameters of the pseudo-magnetic field in the vicinity of the substance surface. Such analysis and new constraints are presented in this work.

The search for the pseudo-magnetic field in the vicinity of the substance surface is motivated by one of the most important questions of the elementary-particle physics – the possibility of axion existence. It is well known that introduction of axion allows to solve the problem of CP-violation in strong interactions. Moreover, axion is considered to be one of the candidates for the dark matter in the Universe. A sufficiently complete overview of both theoretical and experimental parts of the problem is presented in the works [5].

One boson exchange between the scalar ($g_S$) and the pseudo-scalar ($g_P$) vertexes gives the following form of the nucleon interaction potential [6]:

$$V(\mathbf{r}) = \frac{\hbar^2}{8\pi m_n} g_S g_P \mathbf{n} \cdot \boldsymbol{\sigma}_n \left( \frac{1}{\lambda_A r} + \frac{1}{r^2} \right) e^{-r/\lambda_A} \quad (1),$$

where $g_S$ and $g_P$ – dimentionless scalar and pseudo-scalar constants of the nucleon connection with an intermediate boson (axion), $\mathbf{n} = \mathbf{r}/r$ – unit vector between neutron and nucleon, $\lambda_A = \hbar/m_A c$ – typical parameter of the range of forces (Compton wavelength of axion), $m_n$ и $\boldsymbol{\sigma}_n$ – mass and spin of neutron. The correlation ($\boldsymbol{\sigma}_n \cdot \mathbf{n}$) violates P and T invariance.

There are the following equations for the coupling constants $g_S$, $g_P$ and mass of axion [5,6]:

$$g_S = \frac{C_{p,n} m_{n,p}}{f_A} \theta \quad (2),$$

$$g_P = \frac{C_n m_n}{f_A} \quad (3),$$

$$m_A = \frac{z^{1/2}}{1+z} \frac{f_\pi m_\pi}{f_A} \cong \frac{0.6 \text{ meV}}{f_A / 10^{10} \text{ GeV}} \quad (4),$$

where $f_A$ is the energy scale of the breaking Peccei-Quinn symmetry, $\theta$ is model parameter, $C_{p,n}$ are model dependent numerical coefficients, $z = m_u / m_d$, $m_A$ is axion mass defined by relation $m_A f_A \approx m_\pi f_\pi$, where $m_\pi \approx 135$ MeV, $f_\pi \approx 92$ MeV. In our case we will consider $\theta$ as free parameter. Constraints for parameter $\theta$ can be obtained from experiment search for pseudo-magnetic field near the surface of substance. In the end of the article the conclusion about sensitivity of the experiment search for pseudo-magnetic precession for determination of parameter $\theta$ will be discussed.

In the vicinity of the substance surface with a material wall thickness of $d \gg \lambda$ the interaction potential can be described by the following formula [1]:

$$V(z) = \pm \frac{\hbar^2 \lambda_A}{4 m_n} N g_S g_P e^{-z/\lambda} \quad (5),$$

where $N$ – the number of nucleons in a unit of the substance volume, and the sign (+)(-) is determined by the sign of the neutron spin projection onto the substance surface normal. Such a spin-dependent form of the potential is equivalent to an effective magnetic field:

$$H_{SP} = \frac{\hbar^2 \lambda}{4 m_n \mu_n} N g_S g_P e^{-z/\lambda} = 1.7 \cdot 10^{-8} \text{ [Gs} \cdot \text{cm}^2\text{]} \cdot \lambda_A \text{ [cm]} \cdot N \text{ [cm}^{-3}\text{]} \cdot g_S g_P e^{-z/\lambda} \quad (6),$$

where $\mu_n$ – neutron magnetic moment $6 \cdot 10^{-12}$ eV/Gs.

At present there exist astrophysical limits for the axion mass. The unclosed range of masses, or so called "axion-window" is within the range 10 μeV $< m_A <$ 10 meV [5], which corresponds to the parameter $\lambda_A$ within the range 2 cm $> \lambda_A > 2 \cdot 10^{-3}$ cm. Thus the search for the pseudo-magnetic field at short distances from the substance surface is the most actual. There is a sufficiently large number of experiments on the search for the pseudo-magnetic field between nucleons within the range from several meters up to a centimeter, for example [7-11], and only one experiment within the range from several centimeters to a micron [12]. The latest experiment was carried out with the help of UCN, studying quantum states of neutron reflecting from substance in the gravitational field of Earth. Unfortunately, a low statistical accuracy of the experiments with "gravitational" levels does not allow achieving a high accuracy of measurements. The proposal of the work [1] to measure the resonance shift in a magnetic resonance spectrometer with UCN allows to make a significant progress in the measurement accuracy of the pseudo-magnetic field parameters. However, already now, before the realization of this proposal, we can obtain new limits for the parameters of the pseudo-magnetic field from the available data on UCN depolarization during their storage in material traps, for example, from the work [2].

Let us look at the scheme of the experiment on the search for the neutron electric dipole moment [2] (Fig. 1). The traps for UCN storage are located in a weak magnetic field of 0.02 Gs directed vertically. If there are pseudo-magnetic fields in the vicinity of the substance surface then the configuration of the leading magnetic field inside the spectrometer gets more complicated and there occurs an additional neutron spin precession around the pseudo-magnetic field. The pseudo-magnetic neutron spin precession in the vicinity of vertical walls leads to a random neutron spin flip and UCN depolarization during their storage. The pseudo-magnetic neutron spin precession in the vicinity of horizontal walls will cause the neutron resonance shift, if the central and external electrodes are made from materials of different densities. Also the sign of the resonance shift in the upper chamber will be opposite to that in the lower one, if the direction of the guiding magnetic field $H$ changes. The resonance shift effect just corresponds to proposal of the work [1]. However let us return to the UCN depolarization effect.

Let the initial state of the polarization equal to $P_0$ and be directed vertically. After the first collision with a vertical wall there occurs a deviation of polarization from the vertical axis by the angle $\Delta \varphi$, which is determined by the integral value of the pseudo-magnetic field along the surface normal and the time of flight through the pseudo-magnetic field. After averaging by neutron velocity angles with respect to the surface normal we have the following formula:

$$\Delta \varphi = \frac{8 \pi \gamma}{v} \int H_{SP}(z) dz = \frac{2 \pi \hbar^2 \gamma N}{m_n \mu_n v} g_S g_P \lambda_A^2, \quad (7)$$

where $\gamma = 3 \cdot 10^3$ Gs$^{-1}$s$^{-1}$ – gyromagnetic relation, which determines the frequency of the neutron precession in magnetic field, v – neutron velocity.

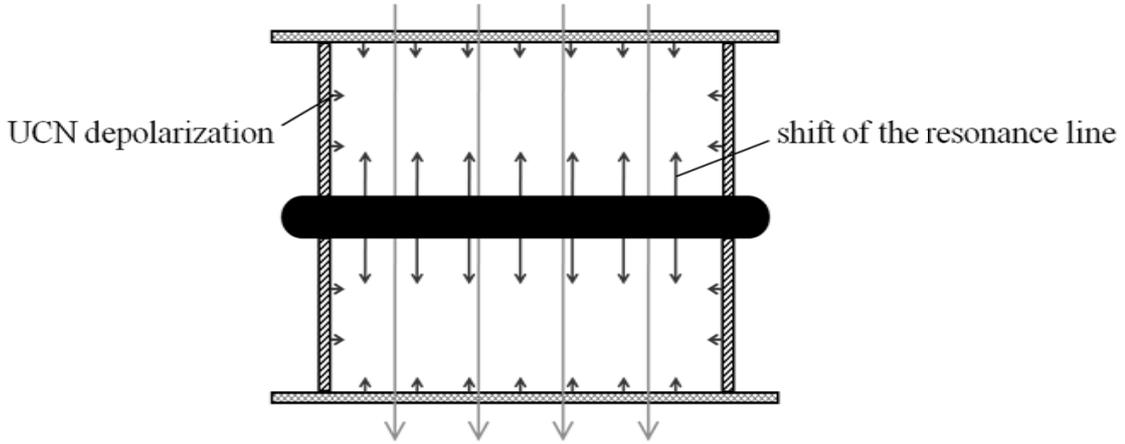

Fig. 1. Scheme of the experiment for neutron electric dipole moment search [2]. The pseudo-magnetic neutron spin precession in the vicinity of vertical walls leads to a random neutron spin flip and UCN depolarization during their storage. The pseudo-magnetic neutron spin precession in the vicinity of horizontal walls will cause the neutron resonance shift, if the central and external electrodes are made from materials of different densities.

We can perform a numerical estimate of the coefficient in the formula (7) for the average UCN velocity v $=5 \cdot 10^2$ cm/s and the density of nucleons in the material of the vertical wall (fused quartz), as in the experimental search for the neutron EDM [2], $N_{SiO_2} = 1.6 \cdot 10^{24}$ cm$^{-3}$.

$$\Delta\varphi \, [\text{rad}] = 4.1 \cdot 10^{18} g_S g_P \lambda_A^2 \, [\text{cm}^2] \quad (8)$$

From this estimate one may conclude that storage of polarized UCN would be impossible at $g_S g_P \lambda_A^2 > 10^{-19}$, because depolarization would occur after the very first collision with the vertical wall.

Let us perform a more accurate estimate for the parameter $g_S g_P \lambda_A^2$ basing on the assumption that UCN polarization in the experiments [2, 3] decreased not more than by 10% after holding UCN in the traps for 100 s.

It should be noted that the frequency of the neutron spin precession in the leading magnetic field of the spectrometer (0.02 Gs) is rather low (60 Hz), and that the neutron spin performs a complete turnover of the precession in the leading magnetic field after having flight 10 cm. Thus the process of emergence of the perpendicular pseudo-magnetic field in the vicinity of the wall at distances shorter than 1 cm should be considered as a non-adiabatic process.

At every new collision with the vertical wall the direction of the neutron spin deviation will be random and uncorrelated to the directions of previous deviations. The most probable full angle of the spin deviation $\varphi$ from the vertical after $n$ collisions will be proportional to the root of the number of collisions $\varphi = \sqrt{n}\Delta\varphi$ or $\varphi = \sqrt{fT}\Delta\varphi$, where $f$ – the frequency of collisions with the vertical wall in a unit of time, and $T$ – UCN storage time. The frequency of collisions with the vertical wall can be calculated via the relation of the vertical surface ($S_V$), the trap volume ($V$), and the neutron velocity (v), $f(\text{v}) = S_V \text{v}/4V$. Although the full angle of the polarization deviation grows with time as a square root function, polarization decreases from its initial state as a linear function, because:

$$P(T) = P_0 \cos\varphi \approx P_0\left(1 - f\frac{\Delta\varphi^2}{2}T\right). \quad (9)$$

Averaging by UCN spectrum and using the parameters of the trap in the experiment [2] and the fact that the decrease in polarization accounted for not more than 10% after storage of UCN for 100 s we can perform an estimation for the value $g_S g_P \lambda_A^2$. Apparently, the UCN polarization observed in the experiment [2] is most likely connected with the presence of paramagnetic centers at the surface of substance and with non-homogeneity of the leading magnetic field, therefore it can be considered as an upper limit for the pseudo-magnetic depolarization.

$g_S g_P \lambda_A^2 \leq 2.96 \cdot 10^{-21}$ [cm$^2$]   (95% C.L.)   (10)

$10^{-4}$ cm $< \lambda_A <$ 1 cm

The upper limit for $\lambda_A$ is determined by the fact that the thickness of the quartz wall in the experiment [2] was 1 cm. The lower limit is determined by the classical method of solving the problem of UCN interaction with the pseudo-magnetic field during UCN reflection, which is true for distances much longer than the neutron wavelength.

Another independent estimation for $g_S g_P \lambda_A^2$ can be made using data from the work [4], where UCN depolarization on different materials was studied. 14 measurements in total were carried out. Beryllium coatings on copper and aluminum, beryllium oxide coatings (i.e. the ones that were used in the traps of the EDM spectrometer [2]) were studied. Moreover depolarization was measured on teflon, fomblin oil (where hydrogen is replaced with fluorine), and also on graphite and graphite coatings. In all cases depolarization was observed. The probability of a spin flip per one reflection from the surface substance ($\alpha$) varies from $(0.6 \div 0.7) \cdot 10^{-5}$ (for teflon, fomblin, graphite and copper) up to $3.75 \cdot 10^{-5}$ (for beryllium oxide coatings).

Results of UCN depolarization measurement in the works [4] and [2] do not contradict each other within coefficient 1.5. The effects of depolarization on copper are lower than those on beryllium coatings. Besides, the density of copper is higher than that of fused quartz by 3.3 times. Therefore the limit for the value $g_S g_P \lambda_A^2$ from the work [4] for Cu is almost by an order of magnitude better.

$g_S g_P \lambda_A^2 \leq 3.9 \cdot 10^{-22}$ [cm$^2$]   (95% C.L.)   (11)

$10^{-4}$ cm $< \lambda_A < 10^{-3}$ cm

The range of values $\lambda_A$ for the given limit if significantly lower, because measurements of depolarization effects were carried out in the magnetic fields of the order of 100 Gs, where the interval of the neutron spin precession accounts for $\sim 1.7 \cdot 10^{-3}$ cm. Accordingly the area of the non-adiabatic precession of the neutron spin in the vicinity of the substance surface is less than $10^{-3}$ cm.

The limits for $g_S g_P \lambda_A^2$ given here are shown in Fig. 2 together with the limits from the works [12] and [10]. Improvement of the limit for $g_S g_P$ in the area of $\lambda_A$ from 0.1 cm to 1 cm accounts for 4÷5 orders of magnitude compared [12].

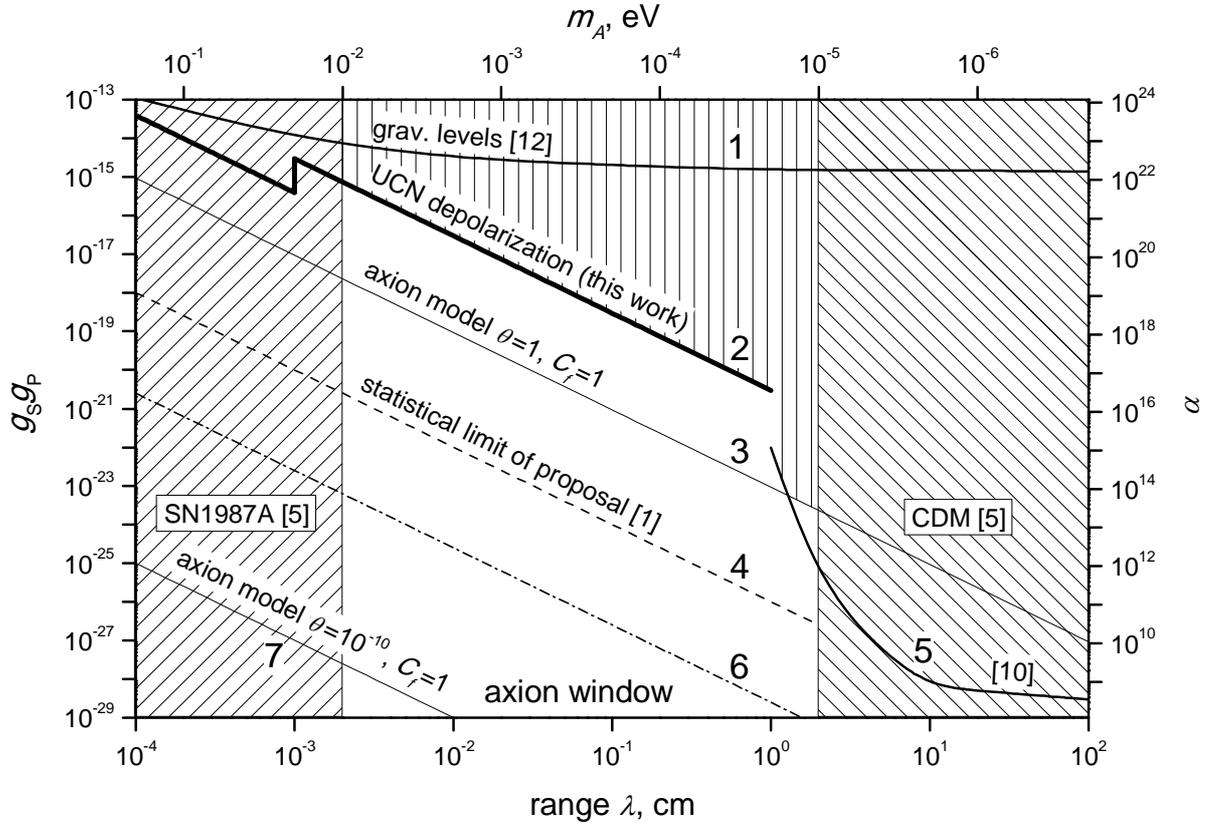

Рис. 2. 1 – "gravitational" levels [12]; 2 – UCN depolarization (this work); 3 – axion model with $\theta=1, C_f=1$; 4 – statistical limit of proposal [1]; 5 – [10]; 6 – constraint for $g_S g_P$ from independent constraints for $g_S$ [15] and $g_P$ [5]; 7 – axion model with $\theta=10^{-10}, C_f=1$.

Further improvement of the accuracy by the method of UCN depolarization measurement, using pure materials not containing paramagnetic and ferromagnetic inclusions is feasible. Particularly employment of fomblin coating in low magnetic fields and cylindrical geometry but with a high vertical wall may possibly allow to make a progress in improving the limit for $g_S g_P \lambda_A^2$ but not more an order of magnitude. It should be mentioned that the sensitivity of experimental search for pseudo-magnetic field by means UCN depolarization is proportional to $g_S g_P \lambda_A^2$ in square, but by means shift of resonance [1] is proportional to $g_S g_P \lambda_A^2$ in power one.

The method of measuring the resonance shift due to the pseudo-magnetic precession is more promising. Of course, simple change of the magnetic field direction in the magnetic shield using a two-chamber trap (Fig. 1) is problematic due to the hysteresis of the magnetic shield. This method will require rotating or moving traps in a magnetic field stabilized by means of Cs-magnetometers. The answer for the question about possible systematic errors can be obtained only by experimental way. The limit of accuracy of the method defined by statistics (UCN density) is shown in Fig. 2 by line 4.

**Discussion**

Now let's make the estimations of the parameters of axion model. Using equations [2-4] one can show that the product $g_S g_P \lambda_A^2 \approx 10^{-23} C_f^2 \theta$, where $C_f^2 = C_p C_n$. As it follows from experimental restrictions for neutron electric dipole moment [2,3] $\theta_{qcd} < 10^{-10}$. Thus the constraint from neutron EDM gives very strong limit, $g_S g_P \lambda_A^2 < 10^{-33}$. The pseudo-magnetic forces at the distance more than $10^{-4}$ cm will be weaker than gravitation forces [13,14]. For comparison with gravitational interaction the right axis in Fig. 2 shows $\alpha = \dfrac{\hbar c}{4\pi G_N m_{n,p}^2} g_S g_P = 1.2 \cdot 10^{37} g_S g_P$, where $G_N$ is Newtonian constant. This analysis shows that in frame of axion model with restriction from neutron EDM $g_S g_P \lambda_A^2$ is so small that it is impossible to reach the comparable restriction by means experimental search for CP-violating pseudo-magnetic field. In Fig. 2 line 4 for the statistical limit of this experiment is considerably higher than line 7 with $\theta_{qcd} < 10^{-10}$. However this conclusion is model dependent. This value $\theta_{qcd} < 10^{-10}$ demonstrates restriction for CP-violation in strong interaction.

It is worth reminding that the limit on $\theta_{qcd}$ of $10^{-10}$ is derived at subatomic distance scales of $10^{-13}$ cm. Staying on pure phenomenological grounds, one could question whether the same stringent limit would apply to $\theta$ measured at distance scales of ~1 cm. Even though we do not have any concrete theoretical realization of scale-dependent $\theta$, it is still worth contemplating such possibility. For example, we can assume that for our range of $\lambda$ ($10^{-4}$ cm ÷ 1 cm) $\theta$ is about one. In general case this assumption is not in contradiction with upper limit for neutron EDM.

Now let us consider constraints for $g_S g_P \lambda_A^2$, using independent limits for $g_S$ and $g_P$. The limits for $g_P$ can be obtained from astrophysical limit for $f_A$, $f_A > 4 \cdot 10^8$ eV [5]. Then $g_P < 2.5 \cdot 10^{-9}$. The constraints for $g_S^2 \lambda_A^2$ in the range $10^{-2}$ cm – 1 cm can be taken from the laboratory experiments for gravitational interaction of material mass [15], $g_S^2 \lambda_A^2 \leq 10^{-40}$. As result we have the following estimation:

$$g_S g_P \lambda_A^2 \leq 2.5 \cdot 10^{-29} \text{ [см}^2\text{]} \qquad (12)$$

This restriction (12) is less than the limit of the experimental sensitivity (line 4). However, we cannot make conclusion that experiment search for CP-violating precession is unreasonable to realize. It would be again model dependent conclusion. In general case the effect of CP-violating interaction of polarized neutron can exist and at the same time effect of interaction of unpolarized matter would be absent due to absence of polarization (attractive forces for one spin direction and repulsive forces for opposite spin direction compensate each other).

**Summary**

1. In frame of general axion model we can assume that $\theta$ is about one for the mass range within "axion window".
2. In general case the considerable pseudo-magnetic field ($g_S g_P \lambda_A^2 > 10^{-26}$ cm$^2$) is possible and small neutron EDM at the same time.

Finally we have to conclude that both experimental tasks: search for CP-violating pseudo-magnetic field and search for neutron EDM should be realized, though the latter is much more important.

Author would like to thank O. Zimmer, M.I. Vysotsky, M. Pospelov, Z. Berezhiani, B.O. Kerbikov and O.M. Zherebtsov for useful discussions and critical remarks. The work was carried out with the support of grant RFBR 07-02-00859.

**References**

[1] O. Zimmer, arXiv:0810.3215v1.
[2] I.S. Altarev et. al., Phys. At. Nucl. 59 (1996) 1152.
[3] C.A. Baker et al., Phys. Rev. Lett. 97 (2006) 131801.
[4] A.P. Serebrov et al., Phys. Lett. A (2003) 373.
[5] C. Amsler et al., Phys. Lett. B 667 (2008) 459.
   G. Raffelt, Lect. Notes Phys. 741 (2008) 51.
[6] J.E. Moody et al., Phys. Rev. D 30 (1984) 130.
[7] D.J. Wineland et al., Phys. Rev. Lett. 67 (1991) 1735.
[8] B.J. Venema et al., Phys.Rev. Lett. 68 (1992) 135.
[9] R.C. Ritter et al., Phys. Rev. Lett. 70 (1993) 701.
[10] A.N. Youdin et al., Phys. Rev. Lett. 77 (1996) 2170.
[11] Wei-Tou Ni et al., Phys. Rev. Lett. 82 (1999) 2439.
[12] S. Baeßler et al., Phys. Rev. D 75 (2007) 075006.
[13] R. Barbieri et al., Phys. Lett. B 387 (1996) 310.
[14] M. Pospelov, Phys. Rev. D 58 (1998) 097703.
[15] J. Chiaverini et al., Phys. Rev. Lett. 90 (2003) 151101.